\documentclass[twocolumn,tighten]{aastex7}

\usepackage{color}
\usepackage{amsmath}
\usepackage{mathtools}

\begin{document}

\title{Molecular and Atomic Gas Associated with the Gamma Ray Supernova Remnant RCW~103}

\author[orcid=0009-0003-8705-5695]{Haruto Inoue}
\affiliation{Faculty of Engineering, Gifu University, 1-1 Yanagido, Gifu 501-1193, Japan}
\email{haruharuinoue131214@gmail.com}  

\author[orcid=0009-0004-3558-3477]{Y. Asano}
\affiliation{Faculty of Engineering, Gifu University, 1-1 Yanagido, Gifu 501-1193, Japan}
\email{z3033004@edu.gifu-u.ac.jp}

\author[orcid=0000-0002-0160-8865]{R. G. Bhuvana}
\affiliation{Faculty of Engineering, Gifu University, 1-1 Yanagido, Gifu 501-1193, Japan}
\email{bhuvanahebbar@gmail.com}

\author[orcid=0000-0001-5609-7372, gname=Rami, sname=Alsaberi]{R. Z. E. Alsaberi}
\affiliation{Faculty of Engineering, Gifu University, 1-1 Yanagido, Gifu 501-1193, Japan}
\affiliation{Western Sydney University, Locked Bag 1797, Penrith South DC, NSW 1797, Australia}
\email{ramy_z@yahoo.com}

\author[orcid=0000-0002-1865-4729]{R. I. Yamada}
\affiliation{Faculty of Engineering, Gifu University, 1-1 Yanagido, Gifu 501-1193, Japan}
\affiliation{Nobeyama Radio Observatory, National Astronomical Observatory of Japan (NAOJ), National Institutes of Natural Sciences (NINS), 462-2 Nobeyama, Minamimaki, Minamisaku, Nagano 384-1305, Japan}
\email{rin.yamada@nao.ac.jp}

\author[orcid=0000-0002-2794-4840]{K. Tsuge}
\affiliation{Faculty of Engineering, Gifu University, 1-1 Yanagido, Gifu 501-1193, Japan}
\affiliation{Nobeyama Radio Observatory, National Astronomical Observatory of Japan (NAOJ), National Institutes of Natural Sciences (NINS), 462-2 Nobeyama, Minamimaki, Minamisaku, Nagano 384-1305, Japan}
\affiliation{Department of Physics, Nagoya University, Furo-cho, Chikusa-ku, Nagoya 464-8601, Japan}
\email{tsuge.kisetsu.i2@f.gifu-u.ac.jp}

\author[orcid=0000-0002-9552-3570]{T. Murase}
\affiliation{Faculty of Engineering, Gifu University, 1-1 Yanagido, Gifu 501-1193, Japan}
\affiliation{Center for Space Research and Utilization Promotion (c-SRUP), Gifu University, 1-1 Yanagido, Gifu, Gifu 501-1193, Japan}
\email{takerun.charvel@gmail.com}

\author[orcid=0000-0002-8966-9856]{Y. Fukui} 
\affiliation{Faculty of Engineering, Gifu University, 1-1 Yanagido, Gifu 501-1193, Japan}
\affiliation{Department of Physics, Nagoya University, Furo-cho, Chikusa-ku, Nagoya 464-8601, Japan}
\email{fukui@a.phys.nagoya-u.ac.jp}

\author[gname=Estela, sname=Reynoso]{E. M. Reynoso} 
\affiliation{Instituto de Astronom\'{i}a y F\'{i}sica del Espacio (IAFE), CONICET–Universidad de Buenos Aires, Ciudad Universitaria, C1428ZAA, Buenos Aires, Argentina}
\email{ereynoso@iafe.uba.ar}

\author[orcid=0000-0002-1411-5410]{K. Tachihara} 
\affiliation{Department of Physics, Nagoya University, Furo-cho, Chikusa-
ku, Nagoya 464-8601, Japan}
\email{k.tachihara@a.phys.nagoya-u.ac.jp} 

\author[orcid=0003-0002-1604-9127]{N. Izumi} 
\affiliation{Department of Physics, Nagoya University, Furo-cho, Chikusa-ku, Nagoya 464-8601, Japan}
\affiliation{National Astronomical Observatory of Japan, 2-21-1 Osawa, Mitaka, Tokyo 181-8588, Japan}
\email{nizumi923@gmail.com}

\author[orcid=0000-0002-6385-8093]{M. Yamagishi}
\affiliation{Institute of Astronomy, The University of Tokyo, 2-21-1 Osawa, Mitaka, Tokyo 181-0015, Japan}
\email{yamagishi@ioa.s.u-tokyo.ac.jp}

\author[orcid=0000-0002-2026-8157]{K. Furuya}
\affiliation{RIKEN Pioneering Research Institute, 2-1 Hirosawa, Wako-shi, Saitama 351-0198, Japan}
\email{kenji.furuya@riken.jp}

\author[0000-0002-6824-6627]{N. Harada}
\affiliation{National Astronomical Observatory of Japan, National Institutes of Natural Sciences, 2-21-1 Osawa, Mitaka, Tokyo 181-8588, Japan}
\affiliation{Department of Astronomy, School of Science, The Graduate University for Advanced Studies (SOKENDAI), 2-21-1 Osawa, Mitaka, Tokyo, 181-1855, Japan}
\email{nanase.harada@nao.ac.jp}

\author[0000-0002-2062-1600]{K. Tokuda}
\affiliation{Faculty of Education, Kagawa University, Saiwai-cho 1-1, Takamatsu, Kagawa 760-8522, Japan}
\email{tokuda.kazuki@kagawa-u.ac.jp}

\author[orcid=0000-0002-9516-1581]{G. Rowell}
\affiliation{School of Physical Sciences, The University of Adelaide, North Terrace, Adelaide, SA 5005, Australia}
\email{gavin.rowell@adelaide.edu.au}

\author[orcid=0000-0002-4990-9288]{M. D. Filipovi{\'c}}
\affiliation{Western Sydney University, Locked Bag 1797, Penrith South DC, NSW 1797, Australia}
\email{m.filipovic@westernsydney.edu.au}

\author[orcid=0000-0003-2062-5692]{Hidetoshi Sano} 
\affiliation{Faculty of Engineering, Gifu University, 1-1 Yanagido, Gifu 501-1193, Japan}
\affiliation{Center for Space Research and Utilization Promotion (c-SRUP), Gifu University, 1-1 Yanagido, Gifu, Gifu 501-1193, Japan}
\affiliation{Department of Intelligence Science and Engineering, Graduate School of Natural Science and Technology, Gifu University, 1-1 Yanagido, Gifu, 501-1193 Japan}
\affiliation{Graduate School of Science, Nagoya University Southern Observatories, Nagoya University, Furo-cho, Chikusa-ku, Nagoya 464-8601, Japan}
\email[show]{sano.hidetoshi.w4@f.gifu-u.ac.jp} 

\correspondingauthor{Haruto Inoue \& Hidetoshi Sano}

\begin{abstract}
We present a study of the interstellar molecular and atomic gas associated with the gamma-ray supernova remnant (SNR) RCW~103 using the Mopra $^{12}$CO($J$~=~1--0), Atacama Large Millimeter/submillimeter Array $^{13}$CO($J$~=~1--0), and the Australia Telescope Compact Array \& Parkes H{\sc i} data. We find that CO clouds in the velocity range from $-58.7$ to $-43.5$~km~s$^{-1}$ show a clear spatial correspondence with the X-ray shell of RCW~103, particularly extending from the northwestern to the southeastern regions. The position--velocity diagram of the CO emission reveals an expanding gas motion with an expansion velocity of $\Delta V \sim$$7.5$~km~s$^{-1}$, which may have been produced by shock waves and/or stellar winds from the progenitor system. We estimate the total cosmic-ray proton energy to be $\sim$$3.0^{+1.1}_{-0.6} \times 10^{47}$~erg, adopting a total interstellar proton density of $\sim$$810$~cm$^{-3}$. This value is more than an order of magnitude lower than those inferred for other gamma-ray–emitting SNRs of comparable age. Given that a substantial amount of dense gas remains within the interior of RCW~103, where thermal X-ray emission is dominant, our results possibly suggest that a significant fraction of the shock energy may have been converted into thermal energy via gas heating, thereby reducing the efficiency of cosmic-ray acceleration.
\end{abstract}


\section{Introduction} \label{sec:introduction}

The origin of cosmic rays, which are mainly composed of protons, is difficult to determine and has been a longstanding question in astronomy for many years. Galactic supernova remnants (SNRs) are most likely to accelerate cosmic rays below $\sim$10$^{15.5}$~eV (= the ``knee energy'') by diffusive shock acceleration \citep[DSA; e.g.,][]{1978MNRAS.182..147B,1978ApJ...221L..29B}. The supernova origin of cosmic rays can be explained in terms of energy balance without problems. Considering the frequency of supernova explosions (1/30 event year$^{-1}$), the cosmic-ray energy density (1.39~eV~cm$^{-3}$), and the escape timescale of cosmic rays from the Galactic disk ($\sim$$3\times10^{14}$~s), the total energy of cosmic rays, $W_\mathrm{p}$, accelerated in a single supernova explosion is estimated to be about $\sim$$10^{50}$~erg, which is equivalent to $\sim$10\% of the energy of a supernova explosion \citep[e.g.,][]{2011piim.book.....D,2013ASSP...34..221G}. However, there is insufficient observational evidence to support these theoretical predictions.

For the observational verification of $W_{\mathrm{p}}$, it is essential to approach the source of acceleration of cosmic-ray protons, which are the main component of cosmic rays. {It is important to study the interstellar medium (ISM) associated with gamma-ray bright SNRs, because cosmic-ray protons produce hadronic gamma-rays through p-p interactions with interstellar protons and the subsequent decay of neutral pions \citep[e.g.,][]{1994A&A...287..959D,2007Ap&SS.309..365G}.} The flux of hadronic gamma rays is proportional to the target gas density and $W_{\mathrm{p}}$, so accurate quantification of the interstellar proton density is essential. The interstellar proton density has been quantified to be approximately a few cm$^{-3}$ using X-ray and infrared observations {\citep[e.g.,][]{2012A&ARv..20...49V}}. However, recent detailed radio line observations of CO and H{\sc i} have revealed that {the ISM surrounding SNRs} is {broadly distributed along their overall morphology and is highly inhomogeneous, with} density contrasts spanning approximately five orders of magnitude \citep[e.g.,][]{2010ApJ...724...59S,2020ApJ...904L..24S,2012ApJ...746...82F}. Therefore, investigating {the interstellar proton density derived from CO and H{\sc i} observations} is crucial for observationally constraining $W_{\mathrm{p}}$.

Recently, \citet{2021ApJ...923...15S,2021ApJ...919..123S,2022ApJ...933..157S}, \citet{2022ApJ...938...94A}{, and \citet{2026ApJ..1001...19M}} presented an SNR age--$W_{\mathrm{p}}$ relation for {16} gamma-ray SNRs. They found that $W_{\mathrm{p}}$ changes with time according to the SNR's evolutionary stages; young SNRs (age $<6$~kyr) show a positive correlation while middle-aged SNRs (age $>8$~kyr) show a negative correlation. The authors concluded that this trend can be possibly explained as a combination of the age-limited acceleration and energy-dependent diffusion (or escape) of cosmic rays. Additionally, the gas densities were derived through analyses of shocked clouds using CO/H{\sc i} datasets. Although the age--$W_{\mathrm{p}}$ relation is crucial for understanding the mechanisms of cosmic ray acceleration and escape in SNRs, the observed samples are not large enough, especially for younger SNRs that would help fill the gap in the more linear positive correlation.

\begin{figure*}[ht!]
\centering
\includegraphics[width=\textwidth]{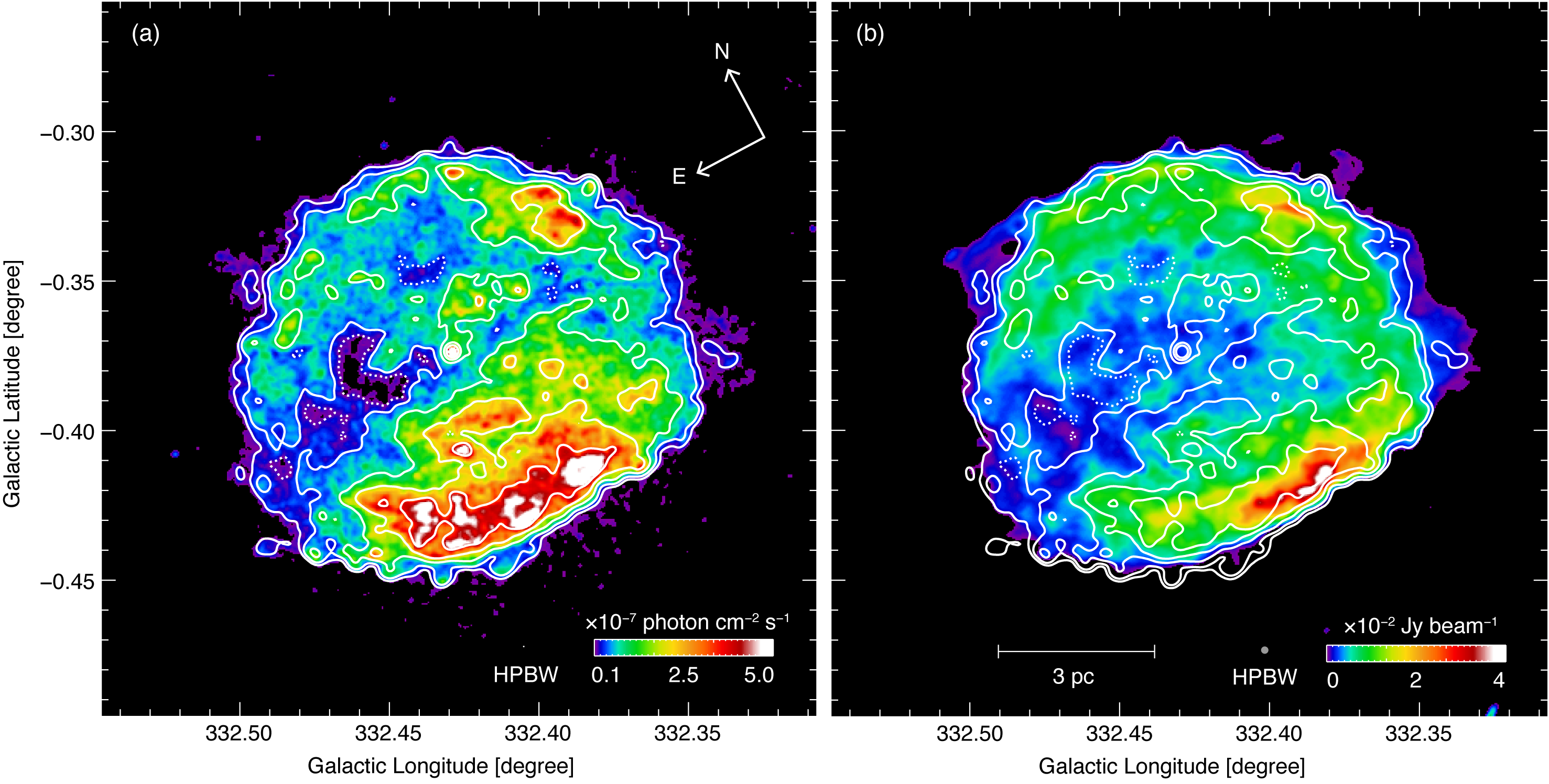}
\caption{Maps of (a) Chandra X-ray flux ($E$:~0.5--7.0~keV) and (b) MeerKAT radio continuum at 1359.7~MHz. The superposed contours are the X-rays whose contour levels are 0.20, 0.40, 1.0, 2.0, and $3.5 \times 10^{-7}$~photon~cm$^{-2}$~s$^{-1}$.}
\label{fig:1}
\vspace*{0.5cm}
\end{figure*}

RCW~103 (also known as G332.4$-$0.4) is a nearly circular Galactic SNR with a diameter of $\sim$10$\arcmin$, exhibitng bright radio continuum emission and thermal-dominated X-ray emission, as shown in Figure~\ref{fig:1}. The low absorbing column density of X-rays and no overlapping sources suggest less contamination along the line of sight \citep[e.g.,][]{2019A&A...629A..51Z}. RCW~103 is classified as a young SNR, with an age estimated to be $\sim$2000--4400~yrs based on the Sedov solution \citep{1997PASP..109..990C,2019MNRAS.489.4444B}. \citet{2024ApJ...970..123X} reported the direction of significant GeV gamma-ray emission from the southern limb of RCW~103 with Fermi-LAT. Given the evidence of interaction with molecular clouds in the same region \citep{1990A&A...240..453O,1993PASA...10..327B,2001AIPC..565..197R}, this gamma-ray emission suggests a possible hadronic origin. 

The kinematic distance of RCW~103 was estimated to be $\sim$3.1~kpc by \citet{2004PASA...21...82R}, based on H{\sc i} absorption observations performed with the Australia Telescope Compact Array (ATCA) and the Parkes radio telescope. Following this, \citet{2006PASA...23...69P} performed $^{12}$CO($J$~=~1--0) line observations at nine points along the southern edge of the remnant. The authors identified a molecular cloud with the intensity peak located at $\sim$$-48$~km~s$^{-1}$, which is consistent with the result of \citet{2004PASA...21...82R}. However, the spatial distribution of molecular clouds across the entire SNR remains unresolved, suggesting that a more detailed investigation of the ISM using CO/H{\sc i} data is necessary.

In the present study, we investigate the ISM associated with RCW~103 by using CO/HI data, combined with X-ray and radio continuum data. The present paper is organized as follows. Section \ref{sec:observations} describes the datasets and Section \ref{sec:result} presents the results, including the distributions and kinematic properties of CO and H{\sc i}. Section \ref{sec:discuss} discusses the ISM associated with the SNR, taking into account its relation with X-ray and gamma-ray emission. Section \ref{sec:conclusion} concludes the paper.

\begin{figure*}[ht!]
\centering
\includegraphics[width=150mm]{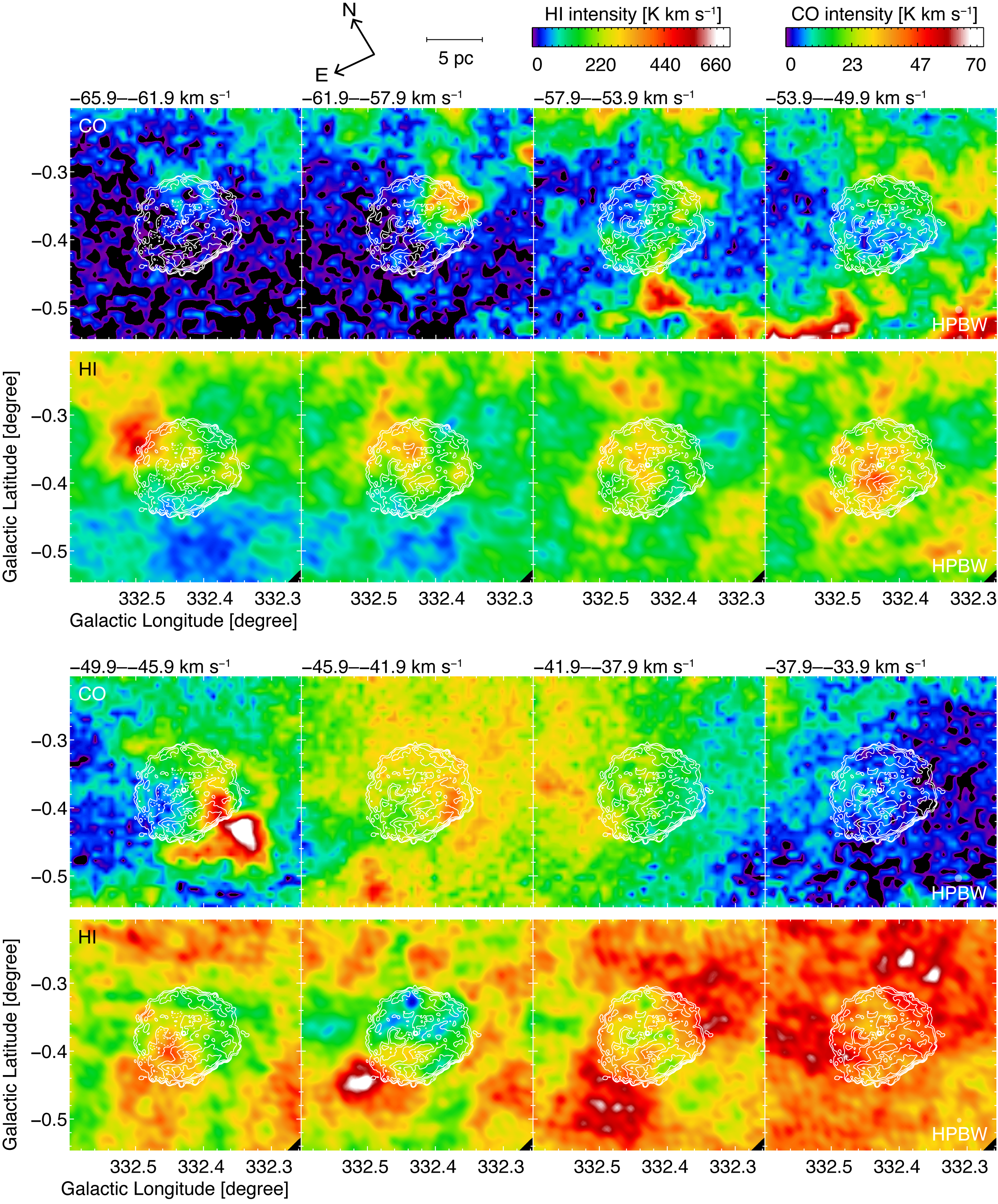}
\caption{Velocity channel maps of the Mopra $^{12}$CO($J$~=~1--0) (upper panels) and the ATCA $\&$ Parkes H{\sc i} (lower panels) data. Each panel of CO/H{\sc i} shows distributions every 4 km s$^{-1}$ in the velocity range from $-$65.9 to $-$33.9~km~s$^{-1}$. The color bars for CO and H{\sc i} are shown at the top of the set of panels. The superposed contours are the same as those shown in Figure \ref{fig:1}. 
\label{fig:2}}
\end{figure*}

\section{Observations} \label{sec:observations}

\subsection{CO} \label{subsec:CO}
We used the archival data of $^{12}$CO($J$~=~1--0) line emission with a central frequency of 115.271~GHz taken with the Mopra 22-m single-dish radio telescope. The beam size was 36$\arcsec$. The typical noise level is $\sim$0.18~K at velocity resolution of 1~km~s$^{-1}$. Details on the observing techniques and data processing can be found in The Mopra Southern Galactic Plane CO Survey Data Release~4 \citep[DR4;][]{2023PASA...40...47C}.\

Observation of $^{13}$CO($J$~=~1--0) emission line were carried out using the Atacama Compact Array (ACA) that is a part of the Atacama Large Millimeter/submillimeter Array (ALMA). We obtained the data using the Band~3 receiver in Cycle~9 (PI: Hidetoshi Sano, \#2022.1.00290.S). We used 8--11 antennas of the 7-m array in 2023 January 18, 20, and 30, and March 5. The mosaic mapping mode was used to cover two regions in the southern part of RCW~103: one centered at $(l,b)=(332\fdg380, -0\fdg422)$ with an area of $2\farcm8 \times 1\farcm3$, and the other centered at $(l,b)=(332\fdg334, -0\fdg437)$ with an area of $1'\times 1'$. The baseline lengths of the 7-m array ranged from 9 to 48~m, corresponding to 3.3--17.6~k$\lambda$ at 110.219~GHz. This bandpass and flux calibrations were performed using J1427$-$4206, and the phase calibration was carried out using J1604$-$4441.

Data calibration and imaging were conducted using the Common Astronomy Software Applications package \citep[CASA, version 5.4.1;][]{2022PASP..134k4501C}. Imaging was performed with the tclean task, applying the multiscale CLEAN algorithm and natural weighting \citep{2008ISTSP...2..793C}. The cleaned 7-m array data were then combined with the calibrated total power (TP) data using the feather task. The synthesized beam size was $15\farcs9\times10\farcs5$ with a position angle of $287\fdg9$, and the typical rms noise was $\sim$0.02~K at a velocity resolution of 0.4~km~s$^{-1}$.

\begin{figure*}[ht!]
\centering
\includegraphics[width=\textwidth]{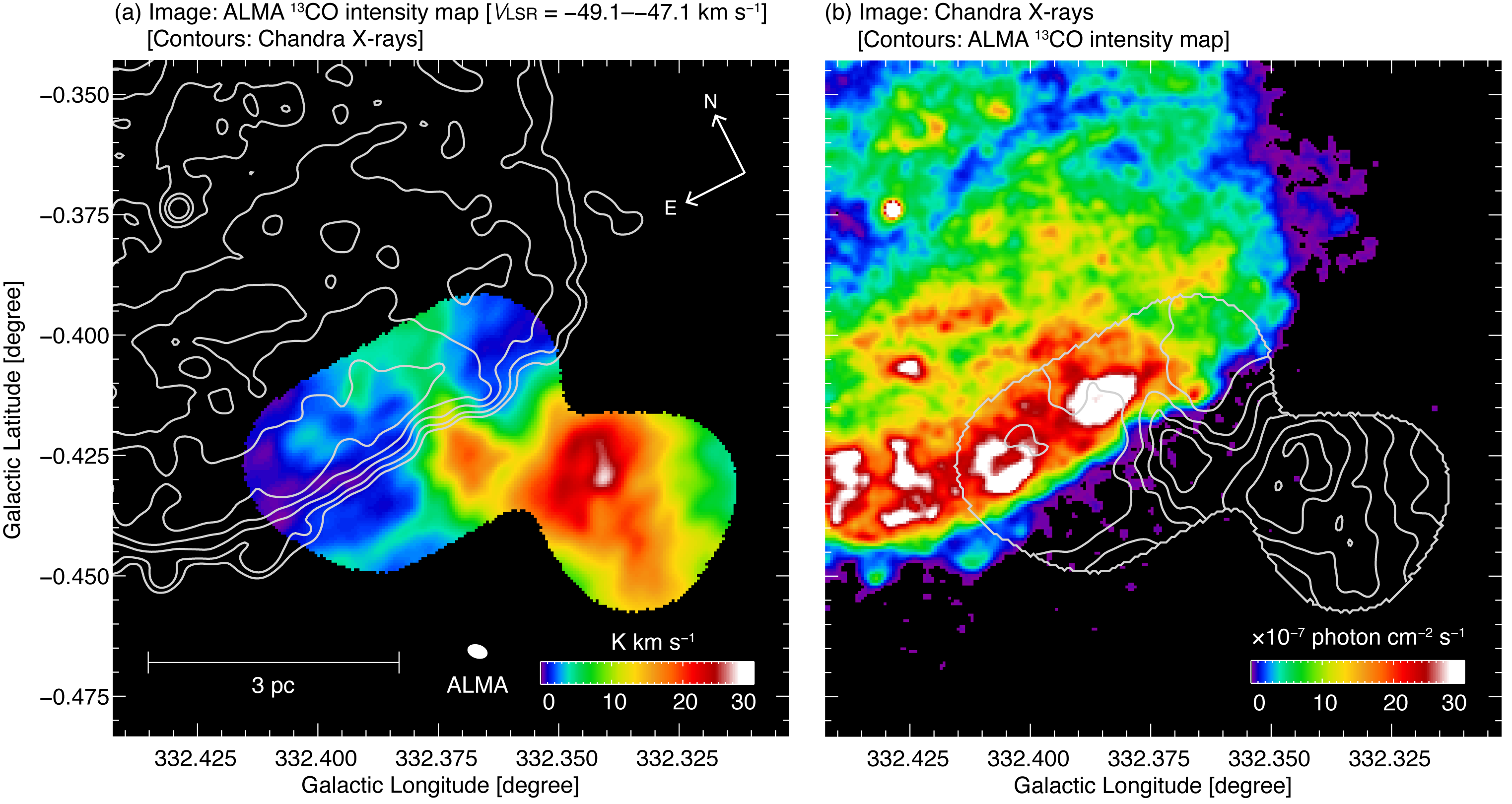}
\caption{(a) Integrated intensity map of ALMA ACA $^{13}$CO($J$~=~1--0), overlaid with Chandra X-ray intensity contours in the energy band 0.5--7.0 keV. The integration velocity range of $^{13}$CO is from $-$49.1 to $-$47.1~km~s$^{-1}$. The contour levels are the same as those shown in Figure \ref{fig:1}. (b) Chandra X-ray intensity map in the 0.5--7.0 keV band, overlaid with contours of the ALMA ACA $^{13}$CO($J$~=~1--0) integrated intensity. The integration velocity range is the same as in \ref{fig:2.5}(a). The contour levels are 3.0, 7.0, 11, 15, 20, and 24 K km s$^{-1}$.  
\label{fig:2.5}}
\vspace*{0.5cm}
\end{figure*}

\begin{figure*}[ht!]
\centering
\includegraphics[width=150mm]{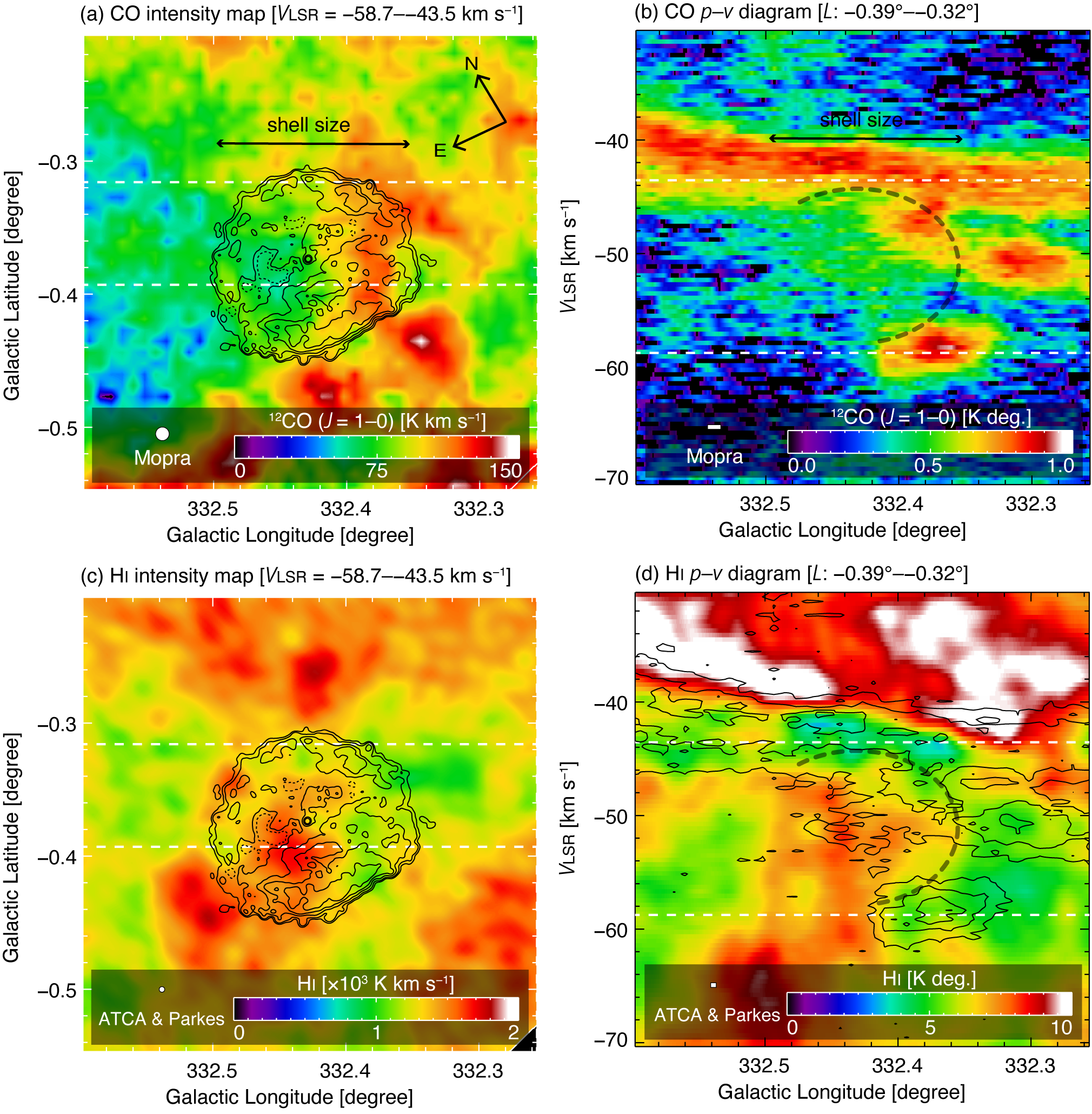}
\caption{Integrated intensity maps and position-velocity (p--v) diagrams of $^{12}$CO($J$~=~1--0) (top panels) and H{\sc i} (lower panels). The integration range is from $-$58.7 to $-$43.5~km~s$^{-1}$ in the velocity for each intensity map; and from $-$0\fdg39 to $-$0\fdg32 in the Galactic longitude for each p--v diagram. Dashed curve delineate cavity-like structure. The superposed contours in Figures~\ref{fig:3}(a) and \ref{fig:3}(c) are the same as those shown in Figure  \ref{fig:1}. The superposed contours in Figure~\ref{fig:3}(d) indicate the $^{12}$CO($J$~=~1--0), whose contour levels are 0.22, 0.44, 0.66, and 0.88 K degree.
\label{fig:3}}
\vspace*{0.5cm}
\end{figure*}

\subsection{H{\sc i}} \label{subsec:HI}

We used the 21~cm H{\sc i} data taken with the ATCA and single-dish data from the Parkes 64-m radio telescope. Details on the observing techniques and data processing can be found in \citet{2004PASA...21...82R}. The beam size of H{\sc i} was 15$\arcsec$. The typical noise level is $\sim$2~K at velocity resolution of 1~km~s$^{-1}$.

\subsection{X-rays} \label{subsec:x-rays}

To compare the spatial distributions of CO/H{\sc i} and X-rays, we used X-ray data {of RCW~103} from the Chandra satellite {\citep[e.g.,][]{2015ApJ...810..113F,2019A&A...629A..51Z}}. These data were obtained from three 
{observations (ObsIDs 970, 11823, and 12224; PI: Gordon Garmire)} with the Advanced CCD Imaging Spectrometer (ACIS), providing a total exposure time of 99.2~ks. The reduction and analysis of the observation data were performed using CIAO 4.5 and CALDB 4.4.10. The CIAO task ``chandra\_repro'' was used with standard parameters to reprocess the data. After that, {a flux image in the 0.5--7.0 keV energy band was created after background subtraction and exposure correction.}

\subsection{Radio Continuum} \label{subsec:radio continuum}

We used the 1359.7~MHz radio continuum data that appear in the SARAO MeerKAT Galactic Plane Survey \citep[SMGPS;][]{2024MNRAS.531..649G}. The angular resolution is $\sim$8$\arcsec$ and the broad-band root-square sensitivity is 10--20~$\text{\textmu}$Jy~beam$^{-1}$.

\section{{Results}} \label{sec:result}

\subsection{Distribution of Radio Continuum and X-Rays} \label{subsec:distribution}

Figure~\ref{fig:1}(a) shows Chandra broadband X-ray image of RCW 103 with contours. It is nearly circular with a diameter of $\sim$$10\arcmin$ and has a filled structure inside. There are two bright regions in the southeast and northwest. The central compact object (CCO) is located at the center of the SNR, and to the northeast, a ``C-shaped'' hole can be seen \citep[e.g.,][]{2014AAS...22335314B, 2019MNRAS.489.4444B}.

Figure~\ref{fig:1}(b) shows the 1359.7~MHz radio continuum image of RCW~103 from MeerKAT, overlaid with X-ray contours. The radio continuum is also circular like the X-ray, with bright regions along the southeastern and northwestern shell edges. However, the region inside the radio shell is relatively faint, and the CCO and ``C-shaped'' hole observed in X-ray are not clearly detected at the radio wavelength. 

\subsection{Velocity Channel Maps of CO and H{\sc i}} \label{subsec:velocity}

Figure~\ref{fig:2} shows integrated velocity channel maps of the Mopra $^{12}$CO($J$~=~1--0) and the ATCA $\&$ Parkes H{\sc i}. We find that the cloud at $V_{\mathrm{LSR}}$~=~$-$61.9$--$45.9~km~s$^{-1}$ exhibits a good spatial correlation with the X-ray shell of RCW~103. This molecular cloud spatially coincides with the X-ray filament in the western part of the SNR shell at $V_{\mathrm{LSR}}$~=~$-$61.9--$-$53.9~km~s$^{-1}$. Similarly, the molecular clouds in the velocity range of $V_{\mathrm{LSR}}$~=~$-$57.9--$-$41.9~km~s$^{-1}$ 
show a spatial correspondence with the southern part of the SNR shell. In particular, the cloud at $V_{\mathrm{LSR}}$~=~$-$49.9--$-$45.9~km~s$^{-1}$ is especially prominent and surrounds the SNR shell. On the other hand, H{\sc i} has no significant spatial correlation with the X-ray shell. The presence of a dip-like feature in the H{\sc i} channel maps inside the SNR shell at $V_{\mathrm{LSR}}$ $\sim$$-45.9$--$-41.9$~km~s$^{-1}$ suggests that foreground H I gas is absorbing the background radio continuum emission from the SNR.


Figure~\ref{fig:2.5} presents a comparison of the integrated intensity map of the ACA
$^{13}$CO($J$~=~1--0) emission with the X-ray intensity distribution in the southern part of the SNR shell. The $^{13}$CO($J$~=~1--0) emission is integrated over the velocity range $V_{\mathrm{LSR}} = -49.1$--$-47.1~\mathrm{km~s^{-1}}$, which shows the best spatial correspondence with the X-ray shell. In particular, a molecular cloud clump located at $(l, b) \sim (332\fdg375, -0\fdg425)$ spatially coincides with a local indentation in the X-ray shell. This velocity range corresponds to a systemic velocity of approximately $-48~\mathrm{km~s^{-1}}$, consistent with previous studies
\citep{2006PASA...23...69P}.




\subsection{Position--Velocity Diagrams}\label{subsec:position}

Hereafter, we focus on the cloud at around $V_{\mathrm{LSR}} = -$58.7--$-$43.5~km~s$^{-1}$ where a significant spatial correspondence with X-ray emission is observed. We perform a comparison with the SNR shell along the velocity axis and identify a cavity-like structure in the cloud at the same velocity range. 


Figures~\ref{fig:3}(a) and \ref{fig:3}(c) show the distributions of CO and H{\sc i} in the velocity range from $-58.7$ to $-43.5~\mathrm{km~s^{-1}}$, superposed on the X-ray emission. The CO emission is distributed along the southeastern part of the SNR shell and extends counterclockwise toward the northern shell, outlining the X-ray shell morphology. In contrast, the H{\sc i} intensity varies by only a factor of about two between the interior and exterior of the shell. In regions where the CO molecular clouds are prominent, the H{\sc i} emission appears relatively weaker.

Figures~\ref{fig:3}(b) and \ref{fig:3}(d) show the position--velocity (p--v) diagrams of CO and H{\sc i}, respectively. The CO p--v diagram reveals a cavity-like structure at Galactic latitudes of $-0\fdg39$ to $-0\fdg32$, which roughly corresponds to the diameter of the SNR. The cavity has a central velocity of $V_{\mathrm{center}} \sim$$51~\mathrm{km~s^{-1}}$ and a velocity width of $\Delta V_{\mathrm{LSR}} \sim$$7.5~\mathrm{km~s^{-1}}$. In contrast, no cavity-like structure is identified in the H{\sc i} p--v diagram. Instead, the H{\sc i} intensity distribution shows an inverse correspondence with the CO emission. The H{\sc i} dip spatially coincident with the CO cloud is generally interpreted as H{\sc i} self-absorption \citep[e.g.,][]{1978AJ.....83.1607S}, and we do not further explore this feature in the present study.




\section{Discussion} \label{sec:discuss}

\subsection{Molecular Clouds Associated with RCW~103} \label{subsec:molecular}

Previous studies have discussed the possibility that molecular clouds at
$V_{\mathrm{LSR}} \sim$$-48$~km~s$^{-1}$ are associated with RCW~103 \citep[e.g.,][]{2004PASA...21...82R,2006PASA...23...69P}. In particular, \citet{2006PASA...23...69P} suggested that a cloud near $\sim$$-48$~km~s$^{-1}$ may be related to the SNR based on $^{12}$CO($J$~=~1--0) and HCO$^{+}$($J$~=~1--0) spectra obtained toward several positions in the southern region. Motivated by these results, we examine our new CO and H{\sc i} analysis over the full extent of the remnant to assess whether the clouds in the velocity range from $-58.7$ to $-43.5~\mathrm{km~s^{-1}}$ are physically associated with RCW~103. In the following, we interpret the spatial and kinematic properties of the gas and discuss the cloud most plausibly related to the SNR.


\subsubsection{Spatial Distribution of the CO clouds} \label{subsubsec:spatial}


We begin by examining the spatial distribution of molecular gas in the velocity range $V_{\mathrm{LSR}}$ $\sim$$-60$--$-40$~km~s$^{-1}$, which shows a clear correspondence with the X-ray shell (see Figure~\ref{fig:2} and Section~\ref{subsec:velocity}). In particular, CO emission is detected toward the western shell at $V_{\mathrm{LSR}} = -61.9$--$-53.9$~km~s$^{-1}$, as well as toward the southern shell at $V_{\mathrm{LSR}} = -49.9$--$-45.9$~km~s$^{-1}$. Furthermore, the ALMA ACA $^{13}$CO($J$~=~1--0) observations show that molecular gas at $V_{\mathrm{LSR}} \sim$$-48$~km~s$^{-1}$ is distributed along the southern portion of the shell (Figure~\ref{fig:2.5}).

Such a spatial correspondence between the molecular gas and the X-ray shell is commonly discussed in the context of shock--cloud interaction in SNRs \citep[e.g.,][]{2017JHEAp..15....1S,2021Ap&SS.366...58S}. According to \cite{2017JHEAp..15....1S}, the thermal X-ray flux is proportional to the column density of the shock-interacting clouds. Given that the X-ray emission from RCW~103 is dominated by thermal components, it is plausible that shock heating of the ambient gas contributes to the observed enhancement of the thermal X-ray emission along the shell.

\subsubsection{Expanding Gas Motion} \label{subsubsec:expanding}
The presence of a cavity-like structure in the CO p--v diagram at $V_{\mathrm{LSR}} = -58.7$--$-43.5$~km~s$^{-1}$ provides additional support for a physical association between the molecular gas and RCW~103 (see Figure~\ref{fig:3}b). Cavity-like features observed toward SNRs are commonly interpreted as signatures of expanding gas motions, which can be produced by supernova shock waves and/or strong stellar winds from the progenitor system (e.g., \citealt{1990ApJ...364..178K,1991ApJ...382..204K}). If the observed structure in RCW~103 reflects such an expansion driven by the SNR shock and/or progenitor, the inferred expansion velocity of $\Delta V \sim$$7.5$~km~s$^{-1}$ is broadly consistent with values previously reported toward other core-collapse SNRs \citep[e.g.,][]{2012ApJ...746...82F,2018ApJ...864..161K,2021ApJ...919..123S,2021ApJ...923...15S}. 

Furthermore, the spatial extent of the expanding gas motion inferred from the p--v diagram is comparable to the diameter of the SNR. This suggests that the SNR shock may have reached the inner wall of a pre-existing cavity created by stellar winds. The shock crossing time is estimated to be $\sim$900~yr, assuming an initial shock velocity of $\sim$5000~km~s$^{-1}$ and a shell radius of $\sim$4.5~pc at the adopted distance of 3.1~kpc {\citep{2004PASA...21...82R}}. This timescale is significantly shorter than the estimated age of RCW~103 {\citep[$\sim$2000--4400~yrs, e.g.,][]{1997PASP..109..990C,2019MNRAS.489.4444B}}
 and is therefore compatible with the proposed scenario.


\begin{figure*}[ht!]
\centering
\includegraphics[width=\textwidth]{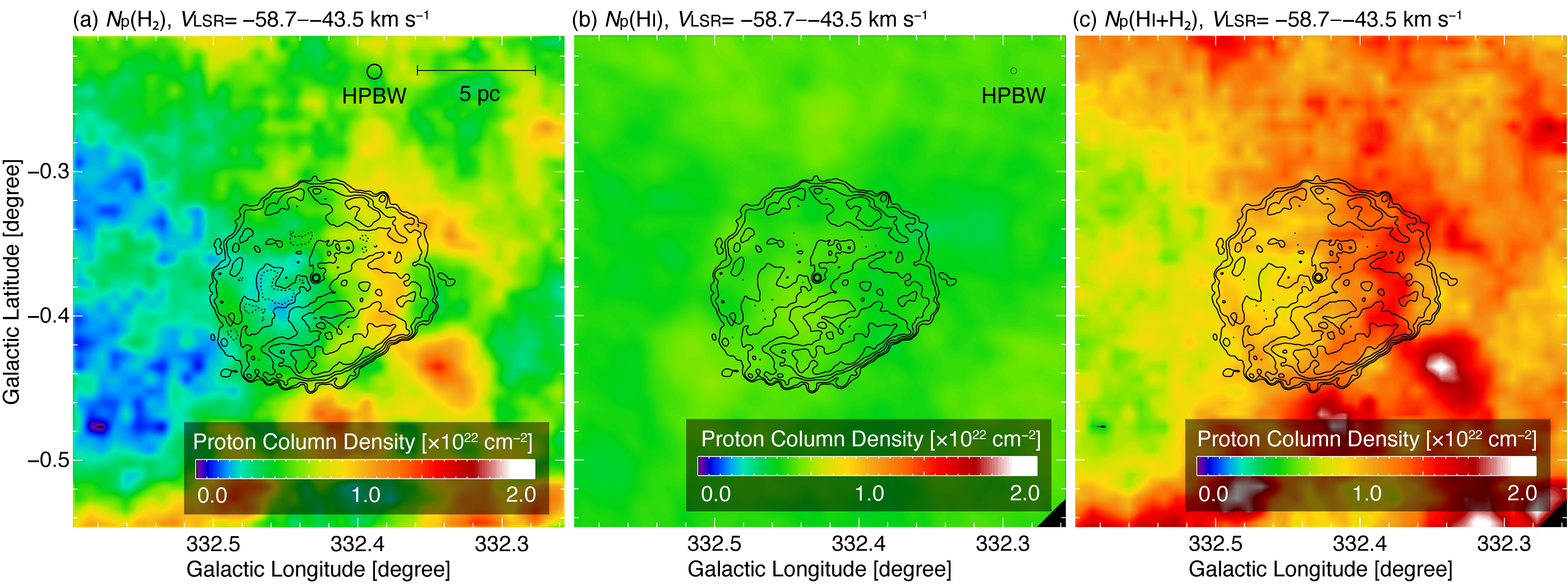}
\caption{Distributions of the ISM proton column densities. (a) $N_{\mathrm{p}}$(H$_2$), (b) $N_{\mathrm{p}}$(H{\sc i}), and (c) $N_{\mathrm{p}}$(H$_2$+H{\sc i}). The velocity range of the integrations is from $-$58.7 to $-$43.5 km s$^{-1}$ for each map. The superposed contours are the same as those shown in Figure~\ref{fig:1}.}
\label{fig:4}
\vspace*{0.5cm}
\end{figure*}

\subsubsection{Comparison with Previous Studies} \label{subsubsec:final}

Finally, we discuss whether the shocked molecular gas at $V_{\mathrm{LSR}} = -58.7$--$-43.5$~km~s$^{-1}$ is consistent with previous observational studies. We first focus on the agreement with earlier radio-line observations. Based on H{\sc i} absorption measurements, \citet{2004PASA...21...82R} reported absorption features toward RCW~103 at $V_{\mathrm{LSR}} \sim$$-43$, $\sim$$-17$, and $\sim$$+1$~km~s$^{-1}$, consistent with the results of \citet{1975A&A....45..239C}. From the presence of the absorption feature at $V_{\mathrm{LSR}} \sim$$-43$~km~s$^{-1}$, these authors estimated a lower limit to the kinematic distance of RCW~103 to be $\sim$3.1~kpc. On the other hand, the systemic velocity inferred from the expanding CO gas motion identified in this study is
$V_{\mathrm{center}} \sim$$-51$~km~s$^{-1}$, which corresponds to a kinematic distance of $\sim$3.5~kpc under the assumption of isotropic expansion. This distance estimate is slightly larger than, but not inconsistent with, the lower limit derived from the H{\sc i} absorption measurements. Therefore, we adopt a distance of 3.5~kpc for the SNR in the subsequent calculations.

It is noteworthy that this result is consistent with recent X-ray proper-motion measurements. Using Chandra observations spanning a 24-yr baseline, \citet{2023ApJ...958...30S} identified inward-moving shock fronts directed toward the center of RCW~103. These inward shocks are most prominent in the southeastern part of the shell. In this region, CO clouds are distributed along the shell, as shown in Figure~\ref{fig:2.5} and Figure~\ref{fig:3}(a), and molecular clumps corresponding to $\sim$0.1~pc-scale indentations seen in the X-ray shell are detected in our ALMA $^{13}$CO observations. In this context, the inward-moving shocks may be interpreted as ``reflected shocks'' generated by the interaction between the SNR shock and dense molecular clouds, providing a coherent explanation for both the X-ray and molecular gas morphologies.

In any case, the molecular gas in the velocity range from $-58.7$ to $-43.5$~km~s$^{-1}$ toward RCW~103 is found to include components that remain intact despite being subject to physical interaction with the SNR.





\subsection{Deriving Total Interstellar Protons}
Having established the molecular and atomic gas associated with RCW~103, we now proceed to quantify their physical properties, such as mass and density. These parameters are required to evaluate the total energy of cosmic-ray protons accelerated by the SNR.

To derive the total interstellar protons, we quantified the proton column densities from the atomic and molecular gas components. The total interstellar proton column density, $N_{\mathrm{p}}$(H{\sc i}+H$_{2}$), is given by the following equation;
\begin{align}
N_{\mathrm{p}}\bigl( \text{H\,\textsc{i}+H}_{2} \bigr)~=~N_{\mathrm{p}}\bigl( \text{H\,\textsc{i}} \bigr)+N_{\mathrm{p}}\bigl( \text{H}_{2} \bigr),
\end{align}
where $N_{\mathrm{p}}$(H$_{2}$) can be derived from the following relation using the $^{12}$CO($J$~=~1--0) integrated intensity $W$(CO);
\begin{align}
N\bigl(\text{H}_{2} \bigr)~=~X_\mathrm{CO} \cdot W\bigl( \text{CO}\bigr),\\
N_{\mathrm{p}}\bigl(\text{H}_{2} \bigr)~=~2 N\bigl(\text{H}_{2}\bigr),
\end{align}
where $X_\mathrm{CO}$ is the CO-to-H$_2$ conversion factor. We used $X_{\mathrm{CO}}~=~0.5\times10^{20}$ cm$^{-2}$~(K~km~s$^{-1}$)$^{-1}$ which was calculated for the region surrounding the SNR RCW~86 near RCW~103 \citep{2019ApJ...876...37S}. 

To estimate the H{\sc i} proton column density $N_{\mathrm{p}}$(H{\sc i}), we adopted a conversion factor between the optical-depth-corrected $N_{\mathrm{p}}$(H{\sc i}) and the H{\sc i} integrated intensity $W$(H{\sc i}), following \citet{2015ApJ...798....6F} and \citet{2017ApJ...850...71F}.
Using this method, the average optical-depth-corrected $N_{\mathrm{p}}$(H{\sc i}) in the RCW~103 region is estimated to be $\sim$$5 \times 10^{21}$~cm$^{-2}$.

Figure~\ref{fig:4} shows the proton column density distributions of $N_{\mathrm{p}}$(H$_2$), $N_{\mathrm{p}}$(H{\sc i}), and their sum, $N_{\mathrm{p}}$(H{\sc i}+H$_2$), integrated over the velocity range from $-58.7$ to $-43.5$~km~s$^{-1}$. While the CO-derived $N_{\mathrm{p}}$(H$_2$) distribution exhibits a clear spatial correspondence with the X-ray emission, the $N_{\mathrm{p}}$(H{\sc i}) distribution appears relatively flat across the region. In the northeastern half of the SNR shell, the values of $N_{\mathrm{p}}$(H$_2$) and $N_{\mathrm{p}}$(H{\sc i}) are comparable, indicating that molecular and atomic gas contribute nearly equally to the total proton content. In contrast, in other regions, particularly toward the southwestern part of the shell, the molecular gas component is dominant. These results indicate that both molecular and atomic gas components must be taken into account when evaluating the interstellar protons associated with RCW~103.




We next derive the masses of the molecular and atomic gas associated with RCW~103.
The molecular gas mass, $M_{\mathrm{H}_{2}}$, and the atomic gas mass, $M_{\mathrm{HI}}$, are calculated as
\begin{eqnarray}
M_{\mathrm{H}_{2}} &=& m_{\mathrm{p}}\,\mu\,\Omega\,D^{2} \sum_{i} N_{i}(\mathrm{H}_{2}), \\
M_{\mathrm{HI}} &=& m_{\mathrm{p}}\,\Omega\,D^{2} \sum_{i} N_{i}(\text{H\textsc{i}}),
\label{eq:mass}
\end{eqnarray}
where $m_{\mathrm{p}}$ is the proton mass, $\mu = 2.8$ is the mean molecular weight, $\Omega$ is the solid angle subtended by each pixel, and $D$ is the distance to the SNR. 

To define the spatial boundary over which the gas mass is integrated, we determine the radius and thickness of the SNR shell based on its radial profile. The radio shell is modeled as a three-dimensional spherical shell with a Gaussian radial profile, from which we derive a shell radius of
$r_{0} = 3.8 \pm 0.7$~pc and a shell thickness of $0.9 \pm 1.6$~pc (see Appendix). Here, the shell thickness is defined as the full width at half maximum (FWHM) of the Gaussian profile,
${\rm FWHM} = 2\sigma\sqrt{2\ln2}$. We estimate the molecular and atomic gas masses associated with the SNR by integrating the column densities within the radial range $r = r_{0} \pm {\rm FWHM}/2$. Using this definition, we obtain $M_{\mathrm{H}_{2}} \sim$$7.2 \times 10^{3}\ M_{\odot}$ and $M_{\mathrm{HI}} \sim$$6.8 \times 10^{3}\ M_{\odot}$.

We then derive the total interstellar proton density, $n_{\mathrm{p}}$, which is given by the sum of the atomic and molecular proton densities, $n_{\mathrm{p}}$(H{\sc i}) and $n_{\mathrm{p}}$(H$_2$). The average number densities of interstellar protons within the radio shell were estimated to be $\sim$420~cm$^{-3}$ for the molecular form $n_{\mathrm{p}}$(H$_{2}$), and $\sim$390~cm$^{-3}$ for the atomic form $n_{\mathrm{p}}$(H{\sc i}) by adopting a shell radius of $\sim$3.8~pc and shell thickness of $\sim$0.9~pc. Then, we finally derived the number density of total interstellar protons $n_{\mathrm{p}}$ to be $\sim$810~cm$^{-3}$.

\begin{figure*}[ht!]
\centering
\includegraphics[width=160mm]{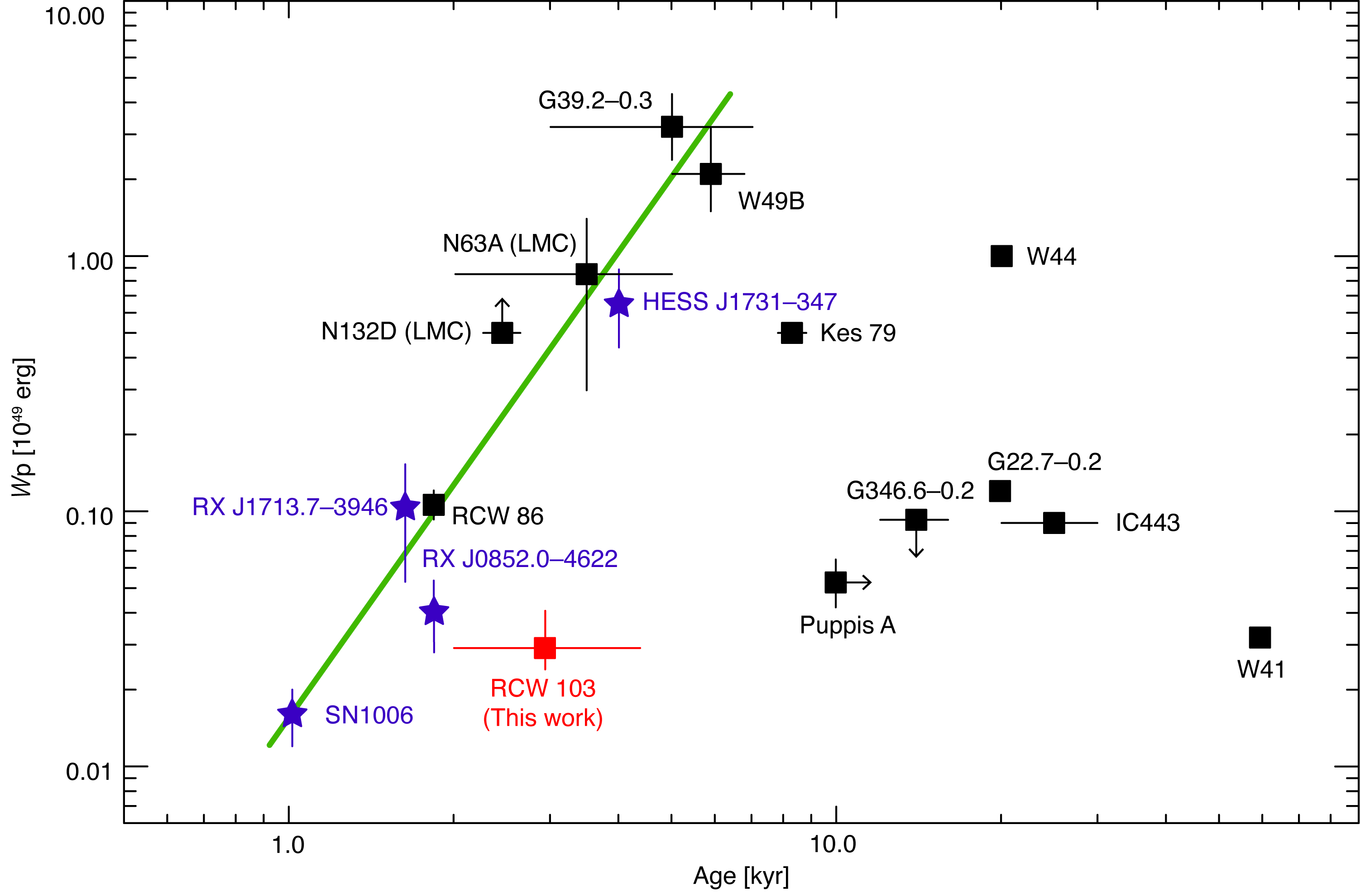}
\caption{Scatter plot between the age of SNRs and the total energy of cosmic-ray protons $W_{\mathrm{p}}$ \citep{2021ApJ...919..123S,2022ApJ...933..157S,2022ApJ...938...94A,2026ApJ..1001...19M}. The green line indicates a least-squares fit to the data points with SNRs ages below 6~kyr. The star symbols denote SNRs whose calculations include the hadronic gamma-ray fraction, while the square symbols denote SNRs calculated under the assumption of a dominant hadronic origin.}
\label{fig:5}
\vspace*{0.5cm}
\end{figure*}

\subsection{Total Energy of Cosmic-Ray Protons} \label{sec:total}

\begin{figure*}[ht!]
\centering
\includegraphics[width=\textwidth]{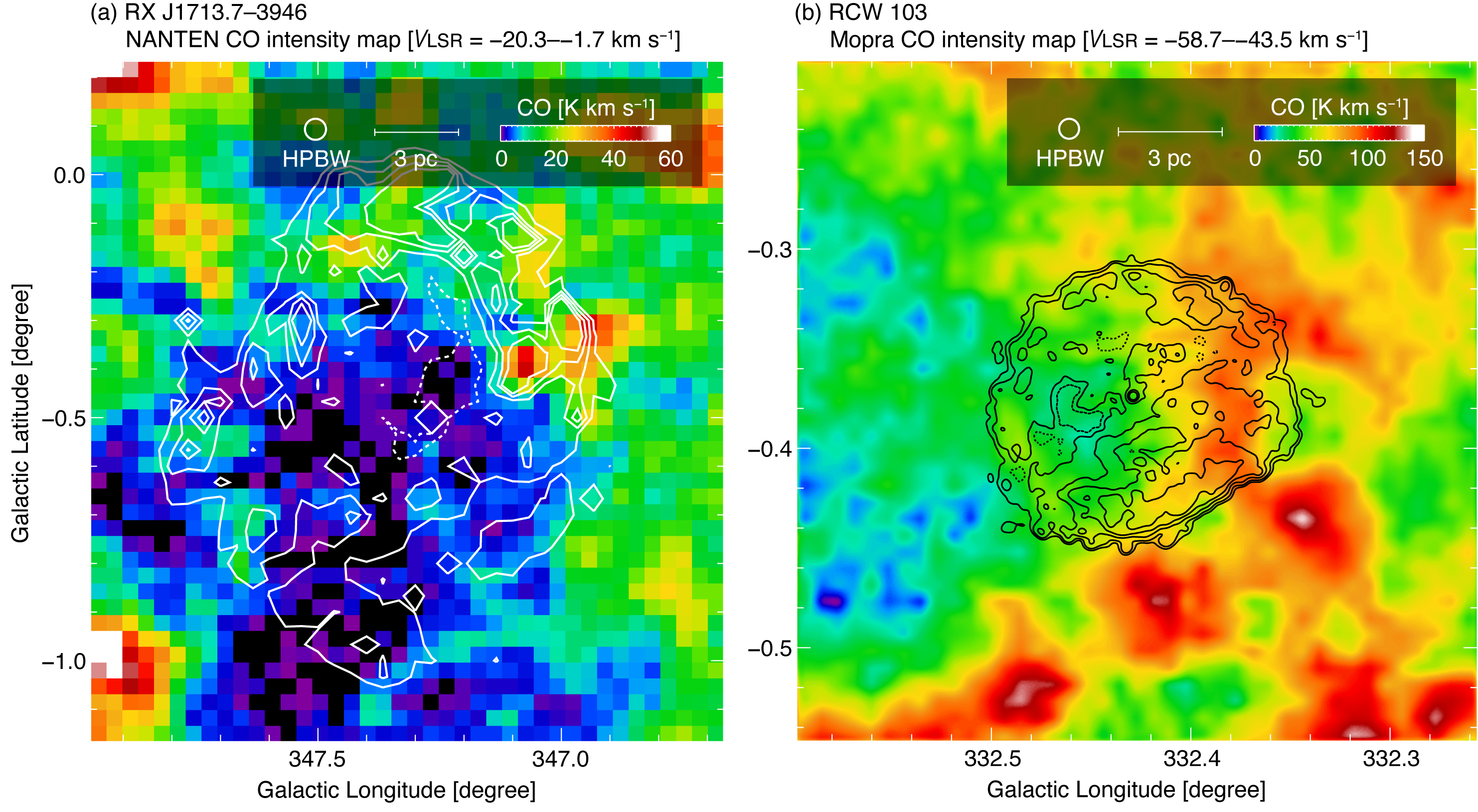}
\caption{(a) Integrated intensity map of NANTEN $^{12}$CO($J$~=~1--0) for RX~J1713.7$-$3946 \citep{2003PASJ...55L..61F}. The superposed contours are the XMM-Newton X-rays, and the contour levels are 30, 57, 85, and 112 photon cm$^{-2}$ s$^{-1}$. (b) Integrated intensity map of Mopra $^{12}$CO($J$~=~1--0) for RCW 103. The superposed contours are the same as those shown in Figure \ref{fig:1}. 
\label{fig:6}}
\vspace*{0.5cm}
\end{figure*}

In many previous studies of gamma-ray emitting SNRs, limited attention has been paid to the detailed distribution of neutral gas. As a result, the ambient gas density is often assumed to be low and spatially uniform, typically of order 1~cm$^{-3}$. Such assumptions, however, are poorly justified and can lead to significant overestimates of the total cosmic-ray energy. In the case of RX~J1713.7$-$3946, \citet{2021ApJ...915...84F} demonstrated that a detailed characterization of the neutral gas distribution is essential for quantifying hadronic gamma-ray emission. In particular, they emphasized the importance of establishing a close spatial correspondence between the distribution of interstellar protons and the observed gamma-ray emission.

An analysis of long-term Fermi-LAT observations presented by \citet{2024ApJ...970..123X} shows that the gamma-ray emission is best reproduced by a point source at the southern edge of RCW~103. Notably, this region overlaps with the molecular gas distribution identified in our study, pointing to a spatial correspondence between the two. Such a correspondence supports the possibility that the observed gamma-ray emission arises from hadronic processes involving interactions between cosmic-ray protons and interstellar protons.


If the gamma-ray emission from RCW~103 is predominantly hadronic in origin, the total energy of cosmic rays can be estimated as follows {\citep{2024ApJ...970..123X}};
\begin{eqnarray}
W_{\mathrm{p}}~=~2.4^{+0.9}_{-0.5} \times 10^{50}\; n_{\mathrm{p}}^{-1}~\mathrm{erg},
\end{eqnarray}
where $n_{\mathrm{p}}$ is the number density of total interstellar protons. We then find $W_{\mathrm{p}}\sim3.0^{+1.1}_{-0.6}\times10^{47}$~erg, by adopting $n_{\mathrm{p}}=810$~cm$^{-3}$.

Figure~\ref{fig:5} presents the relation between the SNR age and $W_{\mathrm{p}}$
\citep{2021ApJ...923...15S,2021ApJ...919..123S,2022ApJ...933..157S,2022ApJ...938...94A,2026ApJ..1001...19M}. RCW~103 appears to be offset by about an order of magnitude below the trend suggested by young SNRs with ages $\lesssim 6$~kyr.



One possible explanation for why $W_{\mathrm{p}}$ in RCW~103 lies about one order of magnitude below the trend seen in previous studies is related to how the shock energy of the SNR is converted. If a substantial fraction of the shock kinetic energy is converted into thermal energy through shock heating of the ambient gas, the energy available for accelerating cosmic rays may be reduced.

Cosmic-ray acceleration is generally expected to be efficient in low-density regions such as inside a wind-blown bubble, where strong shocks can propagate with relatively limited energy loss. By contrast, RCW~103 is characterized by the presence of interstellar gas distributed throughout the interior of the remnant (see Section~\ref{subsec:position}). Moreover, bright thermal X-ray emission is observed across the interior (see Section~\ref{subsec:distribution}), indicating that a substantial fraction of the shock kinetic energy has been converted into thermal energy, heating the gas to temperatures of several tens of millions of kelvin. Taken together, these properties possibly suggest that the dense interior environment of RCW~103 may influence the efficiency of cosmic-ray acceleration.



To examine these possibilities, we compare the X-ray emission and the surrounding ISM environments of RCW~103 with those of the well-studied SNR RX~J1713.7$-$3946 (hereafter refer to as ``RXJ1713''). Figure~\ref{fig:6} presents the CO intensity maps of RXJ1713 and RCW~103. RXJ1713, like RCW~103, is a core-collapse SNR bright in gamma-rays. Recent studies have quantitatively demonstrated that hadronic gamma rays account for approximately 70\% of the total gamma-ray emission from RXJ1713 \citep{2021ApJ...915...84F}. In addition, RXJ1713 has an estimated age of $\sim$1600~yr \citep[e.g.,][]{1997A&A...318L..59W,2016PASJ...68..108T}, comparable to that of RCW~103. Despite these similarities, the dominant X-ray emission mechanisms differ significantly: RXJ1713 is characterized by non-thermal synchrotron X-rays, whereas RCW~103 is dominated by thermal X-rays.

We note that these differences may be related to variations in the gas density within the SNR shell interior: RXJ1713 is associated with dense molecular gas along its shell, while the interior of the remnant is largely evacuated, in contrast to RCW~103. To estimate the gas density within the wind-blown bubble, we derived the column density by integrating over a velocity range selected to exclude foreground and background contributions, and divided it by the shell diameter. The average gas density inside the bubble, $n_{\mathrm{inner}}$, is given by
\begin{equation}
n_{\mathrm{inner}} = N_{\mathrm{inner}} / d,
\end{equation}
where $N_{\mathrm{inner}}$ is the column density derived from the CO intensity in the selected velocity range and $d$ is the shell diameter. Using this approach, we obtain an upper limit of $n_{\mathrm{inner}} \lesssim 5$~cm$^{-3}$ for RXJ1713, whereas the corresponding value for RCW~103 is $\sim$60~cm$^{-3}$. The low gas density inferred for the interior of RXJ1713 is consistent with upper limits derived from thermal X-ray measurements \citep[$n_{\mathrm{inner}} < 2$~cm$^{-3}$, e.g.,][]{2008PASJ...60S.131T}. These results indicate that RCW~103 is filled with relatively dense gas extending into the interior, in clear contrast to RXJ1713.

{In light of the above discussion}, the lower cosmic-ray proton energy inferred for RCW~103 relative to the {16} SNRs examined in previous studies may be related to a larger fraction of the shock energy being converted into thermal energy through gas heating. If this scenario is correct, accurately measuring the distribution and amount of interstellar protons in and around SNRs is essential not only for understanding the mechanisms responsible for thermal and non-thermal X-ray emission, but also for constraining the total energy of accelerated cosmic rays. Testing these ideas will require expanding the sample size. In the era of the Cherenkov Telescope Array \citep[CTA:][]{2019scta.book.....C}, hadronic gamma-ray emission is expected to be detected from many Galactic SNRs that have so far remained undetected at the gamma-ray wavelength, increasing the number of suitable objects for comparative studies by more than an order of magnitude. Furthermore, high-resolution X-ray spectroscopy with ongoing missions such as the X-Ray Imaging and Spectroscopy Mission \citep{2022IJMPD..3130001T} will be crucial for characterizing the thermal plasma properties and elucidating the physical processes of shock heating and particle acceleration in SNRs.




An alternative explanation for the relatively low value of $W_{\mathrm{p}}$ in RCW~103 is that a fraction of the highest-energy cosmic rays has already escaped from the SNR shell. Figure~8 of \citet{2017PASA...34...64L} shows that, within the velocity range of gas associated with RCW~103, a giant molecular cloud is spatially coincident with the TeV gamma-ray source HESS~J1616$-$508. This molecular cloud--gamma-ray complex is located at an offset of $\sim$$+0\fdg3$ in Galactic latitude, corresponding to a projected distance of $\sim$16~pc from the center of the SNR. If, for some reason, the diffusion coefficient around RCW~103 is relatively large, high-energy cosmic rays accelerated in RCW~103 could escape efficiently from the SNR at an early stage and interact with nearby giant molecular clouds, producing gamma-ray emission outside the shell.

If high-energy cosmic rays escape efficiently from young SNRs, particle acceleration at the shock is also expected to be efficient, which would often be accompanied by prominent non-thermal X-ray emission. Therefore, such a cosmic-ray escape scenario alone may not adequately explain why RCW~103, despite its young age, exhibits predominantly thermal X-ray emission. In this aspect, the scenario discussed earlier, in which a substantial fraction of the shock energy is converted into thermal energy through interactions with dense interstellar gas inside the shell, appears more plausible. In either case, further observational studies are required to discriminate between these possibilities. Future observations with facilities such as CTA and ALMA will be essential for investigating the detailed distributions of gamma-ray emission and interstellar protons in and around and inside the SNR RCW~103.

\section{Conclusion} 
\label{sec:conclusion}

We summarize the main results of this study as follows.

\begin{enumerate}
\item 
We investigated the spatial and kinematic distributions of atomic and molecular gas associated with the SNR RCW~103 using the Mopra $^{12}$CO($J$~=~1--0), ALMA $^{13}$CO($J$~=~1--0), and the ATCA \& Parkes H{\sc i} data. The CO clouds in the velocity range from $-58.7$ to $-43.5$~km~s$^{-1}$ show a clear spatial correspondence with the X-ray shell, particularly from the northwestern to the southeastern regions. In contrast, the H{\sc i} gas does not exhibit a pronounced correspondence with the X-ray shell, likely owing to the effects of H{\sc i} self-absorption.
\item 
The CO position--velocity diagram reveals a cavity-like structure, indicating an expanding gas motion driven by shock waves and/or stellar winds from the progenitor system. The inferred expansion velocity is $\sim$7.5~km~s$^{-1}$, with a systemic velocity of $\sim$$-51$~km~s$^{-1}$. This velocity corresponds to a kinematic distance of $\sim$3.5~kpc, providing independent support for distance estimates reported in previous studies.
\item 
The total interstellar proton density was estimated to be $\sim$810~cm$^{-3}$ based on the CO- and H{\sc i}-derived column densities. Using this density, the total energy of cosmic-ray protons was evaluated to be $W_{\mathrm{p}} \sim$$3.0^{+1.1}_{-0.6}\times 10^{47}$~erg. RCW~103 lies approximately one order of magnitude below the trend defined by other gamma-ray SNRs in the age--$W_{\mathrm{p}}$ relation, even when the uncertainty in its age of 2000--4400~yr is taken into account.
\item 
RCW~103 is a young gamma- and X-ray bright SNR with properties broadly comparable to those of RX~J1713.7$-$3946; however, their internal ISM environments differ markedly. While the interior of RX~J1713.7$-$3946 is characterized by significantly low gas density and synchrotron-dominated X-rays, RCW~103 contains relatively dense gas ($\sim$60~cm$^{-3}$) extending throughout its interior and exhibits predominantly thermal X-rays. These differences possibly suggest that, in RCW~103, a substantial fraction of the shock energy may have been converted into thermal energy through gas heating, potentially leading to a lower efficiency of cosmic-ray acceleration.
\end{enumerate}

\begin{figure}[h]
\centering
\includegraphics[width=\columnwidth]{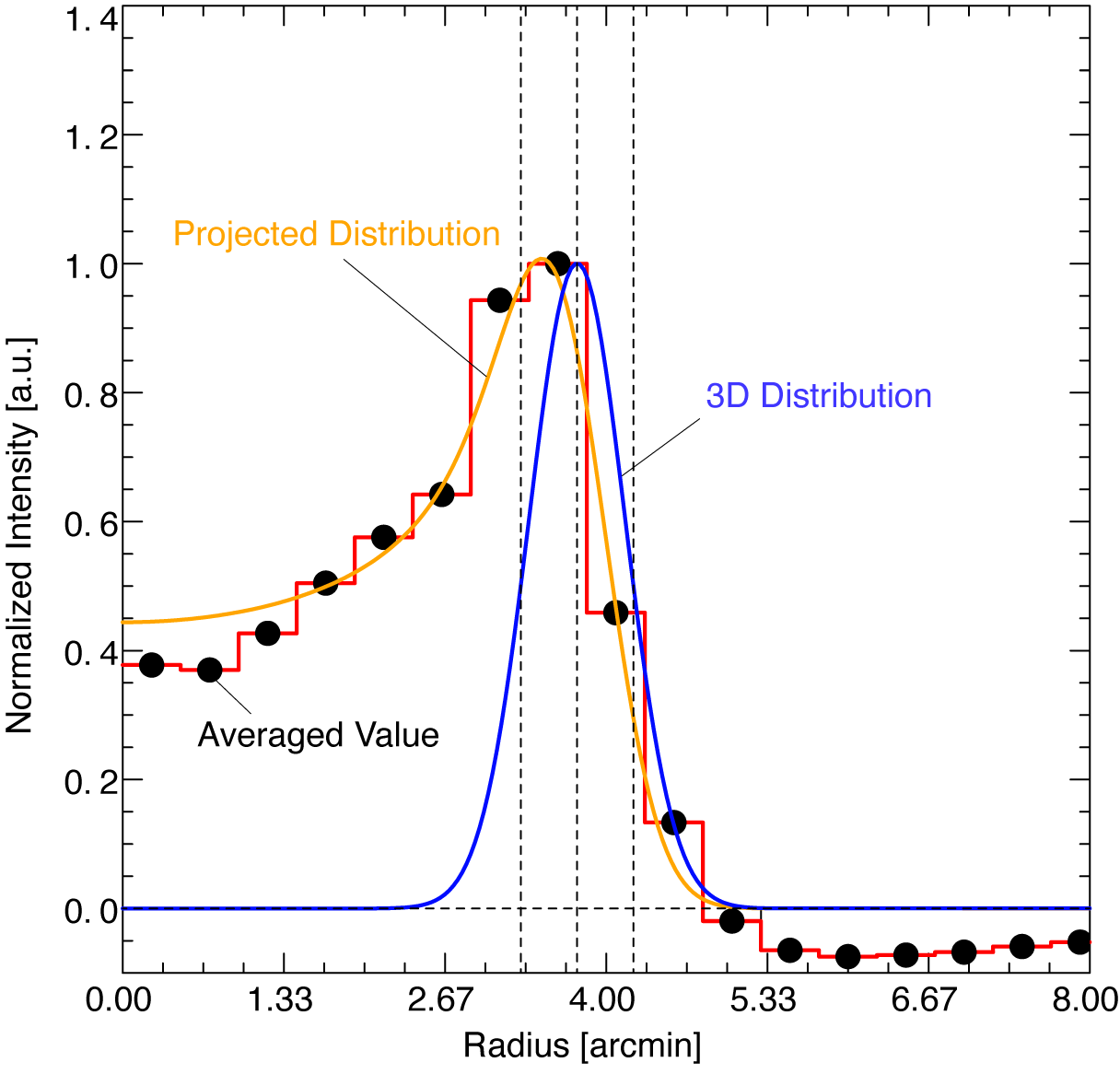}
\caption{Radial profiles of the 1359.7 MHz radio continuum, centered at ($l, b$)~=~(332$\fdg$42, $-$0$\fdg$37). The black filled circles represent the average values of the radio continuum in each annulus. The blue line represents the three-dimensional Gaussian distribution and the orange solid line represents the projected distribution derived from least-squares fitting (see the text for details). The vertical dashed lines indicate the best-fit values for the radius and the range of the shell thickness.
\label{fig:7}}
\end{figure}

\begin{acknowledgments}
We are grateful to Mr. Yoshizumi Okada, Dr. Hiromasa Suzuki, and Dr. Tsunefumi Mizuno for the lively and fruitful discussion on the subject. This paper makes use of the following ALMA data: ADS/JAO.ALMA\#2022.1.00290.S ALMA is a partnership of ESO (representing its member states), NSF (USA) and NINS (Japan), together with NRC (Canada), NSTC and ASIAA (Taiwan), and KASI (Republic of Korea), in cooperation with the Republic of Chile. The Joint ALMA Observatory is operated by ESO, AUI/NRAO and NAOJ. This work also makes use of data obtained from the Chandra Data Archive (Observation IDs 970, 11823, 12224) and software provided by the Chandra X-ray Center in the CIAO package (v4.5). {This research employs a list of Chandra datasets, obtained by the Chandra X-ray Observatory, contained in~\dataset[DOI: 10.25574/cdc.555]{doi.org/10.25574/cdc.555}.} The Mopra telescope is part of the Australia Telescope National Facility (https://ror.org/05qajvd42) which is funded by the Australian Government for operation as a National Facility managed by CSIRO. We acknowledge the Gomeroi people as the Traditional Owners of the Observatory site. The NANTEN project is based on a mutual agreement between Nagoya University and the Carnegie Institution of Washington (CIW). We greatly appreciate the hospitality of all the staff members of the Las Campanas Observatory of CIW. We are thankful to many Japanese public donors and companies who contributed to the realization of the project. This work is financially supported by a grant-in-aid for Scientific Research (KAKENHI, Nos. 20KK0309, 22H00152, ﻿24H00246, and 26K21722) from MEXT (the Ministry of Education, Culture, Sports, Science and Technology of Japan). This work was also supported by NAOJ ALMA Scientific Research Grant Code 2023-25A. Haruto Inoue and Hidetoshi Sano were supported by the ALMA Japan Research Grant of NAOJ ALMA Project, NAOJ-ALMA-405. We thank the anonymous referee for valuable and constructive comments that significantly improved the manuscript.
\end{acknowledgments}





%
\facilities{ALMA, Mopra, CXO, NANTEN, Australia Telescope Compact Array (ATCA), Parkes, MeerKAT}


\software{IDL Astronomy User's Library \citep{1993ASPC...52..246L}, MIRIAD\citep{1995ASPC...77..433S}, CASA\citep[v5.4.1;][]{2022PASP..134k4501C}, CIAO\citep[v4.5;][]{2006SPIE.6270E..1VF},CALDB\citep[v4.4.10;][]{2007ChNew..14...33G}}




\vspace*{0.5cm}
\section*{APPENDIX: radial profile of radio continuum shell}
To determine the apparent diameter and thickness of the radio continuum shell\footnote{{We adopted the radio continuum image for this geometrical analysis because the synchrotron shell provides a relatively continuous tracer of the global SNR boundary, whereas the X-ray surface brightness is more strongly affected by spatial variations in the physical conditions of the hot plasma.}}, we {fitted the} radial profiles {with} a three-dimensional isotropic spherical-shell {model characterized by a Gaussian radial profile $f(r)$, while} shifting the assumed center around the geometric center of the SNR. {We adopted the Gaussian function as a simple empirical approximation, rather than as a physical model of the intrinsic shell emissivity, because it provides an objective and reproducible measure of the shell radius and characteristic width (FWHM) and has been successfully applied to characterize shell-like emission profiles in previous studies \citep[e.g.,][]{2007ApJ...661..236A,2021ApJ...923...15S,2025ApJ...978..123I}.} The function $f(r)$ is described as:
\begin{eqnarray}
f(r) =  A \exp[-(r-r_0\bigr)^2/2\sigma^2]
\label{eqxx}
\end{eqnarray}
where $A$ is a normalization constant of the Gaussian profile, $r_0$ is the shell radius, and $\sigma$ is the standard deviation of the Gaussian profile.

As a result, we obtained the center position of the shell as $(l, b)~=~(332\fdg424, -0\fdg374)$ with the minimum chi-square value of the least-squares fitting. Figure \ref{fig:7} shows the radial profile of the radio continuum, centered at $(l, b)~=~(332\fdg424, -0\fdg374)$. We derived a shell radius $r_{\mathrm{0}}~=~3\farcm76 \pm 0\farcm65$ ($\sim$$3.83\pm0.66$~pc) and a thickness of $0\farcm93 \pm 1\farcm56$ ($\sim$$0.94 \pm 1.59$~pc) as the optimal fitting parameters. Here, the thickness of the shell is defined as the full width at half maximum (FWHM) of the Gaussian function, i.e., $2\sigma\sqrt{2\ln2}$.

\bibliography{01_references}{}
\bibliographystyle{aasjournal}




\end{document}